\documentclass[10pt, conference, letterpaper]{IEEEtran}
\IEEEoverridecommandlockouts
\usepackage[T1]{fontenc}
\usepackage[noend]{algpseudocode}
\usepackage{subcaption}
\usepackage{algorithm}
\usepackage{booktabs}
\usepackage{comment}
\usepackage[draft]{hyperref}
\usepackage{amsmath,amssymb,amsfonts}
\usepackage{graphicx}
\usepackage{booktabs}
\usepackage{mathtools}
\usepackage{textcomp}
\usepackage{xcolor}
\usepackage{float}
\usepackage{bm}
\usepackage{bbold}
\usepackage[normalem]{ulem}
\usepackage[inline]{enumitem}
\usepackage{eso-pic}

\def\BibTeX{{\rm B\kern-.05em{\sc i\kern-.025em b}\kern-.08em
    T\kern-.1667em\lower.7ex\hbox{E}\kern-.125emX}}

\usepackage{lipsum}  

\usepackage[para]{footmisc}

\newif\iftodos
\todostrue

\begin{document}
\title{RouteNet-Erlang: A Graph Neural Network \\for Network Performance Evaluation}
\makeatletter
\newcommand{\newlineauthors}{%
  \end{@IEEEauthorhalign}\hfill\mbox{}\par
  \mbox{}\hfill\begin{@IEEEauthorhalign}
}
\makeatother

\author{
    \IEEEauthorblockN{Miquel Ferriol-Galmés\IEEEauthorrefmark{1}, Krzysztof Rusek\IEEEauthorrefmark{2}, José Suárez-Varela\IEEEauthorrefmark{1}, Shihan Xiao\IEEEauthorrefmark{3}, \\ Xiang Shi\IEEEauthorrefmark{3}, Xiangle Cheng\IEEEauthorrefmark{3}, Bo Wu\IEEEauthorrefmark{3}, Pere Barlet-Ros\IEEEauthorrefmark{1}, Albert Cabellos-Aparicio\IEEEauthorrefmark{1}}
    \IEEEauthorblockA{\IEEEauthorrefmark{1}Barcelona Neural Networking Center, Universitat Politècnica de
Catalunya, Spain\\
 Corresponding email: miquel.ferriol@upc.edu}
    \IEEEauthorblockA{\IEEEauthorrefmark{2}AGH University of Science and Technology, Poland}
    \IEEEauthorblockA{\IEEEauthorrefmark{3}Huawei Technologies Co., Ltd., China}

    \thanks{This publication is part of the Spanish I+D+i project TRAINER-A (ref.PID2020-118011GB-C21), funded by MCIN/ AEI/10.13039/501100011033. This work is also partially funded by the Catalan Institution for Research and Advanced Studies (ICREA) and the Secretariat for Universities and Research of the Ministry of Business and Knowledge of the Government of Catalonia and the European Social Fund.}
\vspace{-0.8cm}
}

\AddToShipoutPictureBG{
  \AtPageLowerLeft{%
    \raisebox{3\baselineskip}{\makebox[\paperwidth]{\begin{minipage}{18cm}
    \footnotesize \textbf{NOTE:} This paper has been accepted for publication at IEEE INFOCOM 2022. ©2022 IEEE. Personal use of this material is permitted. Permission from IEEE must be obtained for all other uses, in any current or future media, including reprinting/republishing this material for advertising or promotional purposes, creating new collective works, for resale or redistribution to servers or lists, or reuse of any copyrighted component of this work in other works.
    \end{minipage}}}%
  }
}

\maketitle

\begin{abstract}
Network modeling is a fundamental tool in network research, design, and operation. Arguably the most popular method for modeling is Queuing Theory (QT). Its main limitation is that it imposes strong assumptions on the packet arrival process, which typically do not hold in real networks. In the field of Deep Learning, Graph Neural Networks (GNN) have emerged as a new technique to build data-driven models that can learn complex and non-linear behavior. In this paper, we present \emph{RouteNet-Erlang}, a pioneering GNN architecture designed to model computer networks. RouteNet-Erlang supports complex traffic models, multi-queue scheduling policies, routing policies and can provide accurate estimates in networks not seen in the training phase. 
We benchmark RouteNet-Erlang against a state-of-the-art QT model, and our results show that it outperforms QT in all the network scenarios.
\end{abstract}



\vspace{0cm}
\section{Introduction}
\vspace{0cm}



Network modeling is central to the field of computer networks. Models are useful in researching new protocols and mechanisms, allowing administrators to estimate their performance before their actual deployment in production networks. Network models also help to find optimal network configurations, without the need to test them in production networks. 

Queuing Theory (QT) \cite{cooper1981queueing} is arguably the most popular modeling technique, where networks are represented as inter-connected queues that are evaluated analytically. This represents a well-established framework that can model complex and large networks. Its main limitation is that it imposes strong assumptions on the packet arrival process, which typically do not hold in real networks~\cite{xu2018experience}. Internet traffic has been extensively analyzed in the past two decades~\cite{traffic1, traffic2, traffic3, traffic4, traffic6} and, despite the community has not agreed on a universal model, there is consensus that in general aggregated traffic shows strong autocorrelation and a heavy-tail~\cite{traffic5}. 



Another network modeling alternative is computational models (e.g., network simulators), which provide excellent accuracy. State-of-the-art network simulators include a wide range of network, transport, and routing protocols, and are able to simulate realistic scenarios. However, this comes at a very high computational cost that depends linearly on the number of packets being simulated. As a result, they are impractical in scenarios with realistic traffic volumes or large topologies. In addition, and because they are computationally expensive, they do not work well in real-time scenarios.



Machine Learning (ML) \cite{mnih2015human} provides a new breed of mechanisms to model complex systems. 
In particular, Deep Learning (DL) \cite{lecun2015deep} 
 has demonstrated to extract deep knowledge from human-understandable descriptions of a system.
This approach has proven to 
achieve unprecedented accuracy in modeling properties of complex systems, like proteins~\cite{AlphaFold2021}.

The main advantage of DL models is that they are \emph{data-driven}. DL models can be trained with real-world data, without making assumptions about the system. This enables to build models with unprecedented accuracy by effectively modeling the entire range of non-linear and multidimensional system characteristics. Computationally, DL is based on linear algebra and can take advantage of massive parallelism leveraging dedicated hardware and compilers.




Within the field of DL, Graph Neural Networks (GNN)~\cite{scarselli2008graph} have recently emerged as an effective technique to model graph-structured data. 
GNNs are tailored to understand the complex relationships between the elements of a graph. The main novelty of GNNs is that their internal architecture is dynamically assembled based on the elements and connections of input graphs, and this enables to learn universal modeling functions that are invariant to graph isomorphism. GNNs are thus able to \emph{generalize} over graphs, which means that they can produce accurate estimates in different graphs not seen during training. As we will show in this paper, this is a critical advantage of GNNs in the context of network modeling.

The novel GNN paradigm finally allows the application of ML in domains where data is essentially represented as graphs. As a consequence, at the time of this writing, substantial research efforts are being devoted to applying GNNs to different fields where data is fundamentally represented as graphs, such as chemistry~\cite{gilmer2017neural}, physics~\cite{battaglia2016interaction} and others~\cite{zhou2018graph}~\cite{Lange2020}.  

We argue that GNNs represent a new network modeling language with attractive advantages and characteristics. GNNs are designed to learn graphs, and computer networks are fundamentally graphs of connected queues. However, GNNs are not a black-box that map data inputs to outputs, it is actually a \emph{modeling tool} that needs to be researched and designed to account for the core behavior of computer networks. In contrast to more classical DL models, where the architecture is basically defined by the number of layers and neurons, GNNs are assembled ad-hoc, based on the elements and connections of the input graphs. 
These components represent the GNN modeling language, and they need to be carefully designed to reflect the relevant properties of the system being modeled.

In this paper, we present RouteNet-Erlang (RouteNet-E), a novel GNN-based architecture designed for performance evaluation of computer networks. RouteNet-E shares the same goals as QT models: it is also able to model a network of queues, with different sizes and scheduling policies, while providing accurate estimates of delay, jitter, and losses. Interestingly, RouteNet-E is not limited to Markovian traffic models as QT, but rather it supports arbitrary traffic models including more complex ones with strong autocorrelation and high variance, which better represent the properties of real-world traffic~\cite{traffic5}. We also show that RouteNet-E overcomes one of the main limitations of existing ML-based models: \textit{generalization}. RouteNet-E is able to make accurate estimates in samples of unseen topologies 
one order of magnitude larger than those seen during training. 

We benchmark RouteNet-E against a state-of-the-art QT model, over a wide variety of network samples covering several different traffic models, from basic Poisson, to more realistic and complex models with strong autocorrelation and approximated heavy-tails. Our evaluation results show that the proposed model outperforms the QT benchmark in \emph{all} the network scenarios evaluated, always producing accurate delay predictions with a worst-case error of $6\%$ (for QT is $68\%$). We also show RouteNet-E's remarkable performance in hundreds of random network topologies not seen during training. Lastly, we measure its inference speed, which is in the order of milliseconds, in line with the QT benchmark.

All datasets, code, and trained models of RouteNet-E used in this paper are publicly available at~\cite{github-erlang}.

\vspace{0cm}
\section{Challenges of GNN-based network modeling} \label{sec:challenges}
\vspace{0cm}

In this section, we describe the main challenges that GNN-based solutions need to address for network modeling. These challenges drove the core design of RouteNet-E.

\textbf{Traffic models:} A model is an abstraction of a system able to distill the essential aspects of the system. Networks transport millions of packets and, as a result, network models require a useful abstraction for packets, that is why supporting arbitrary stochastic traffic models is crucial. 
Experimental observations show that traffic on the Internet has strong autocorrelation and heavy-tails~\cite{traffic5}. In this context, it is well-known that a main limitation of QT is that it fails to provide accurate estimates on realistic Markovian models with continuous state space, or non-Markovian traffic models.
To the best of our knowledge, analytical models for queues with general arrival processes are limited to infinite buffers~\cite{Norros94}, or they make some sort of approximation (e.g., asymptotic), which greatly differs from the actual behavior of computer networks.
The challenge for GNN-based modeling is, how to design an architecture that can accurately model realistic traffic models? 

\textbf{Training and Generalization:} One of the main differences between analytical modeling (e.g., QT) and ML-based modeling is that the latter requires training. In ML, training involves obtaining a \emph{representative} dataset with network measurements. The dataset needs to include a broad spectrum of network operational regimes. In practice, this means testing how different congestion levels affect performance metrics (delay, jitter, and losses), evaluating how different queuing policies affect performance or testing different routing policies, among others. Without this, the ML model is unable to learn and provide accurate estimates. Note that this is a common property of all neural network architectures.
Generating this training dataset from networks in production is typically unfeasible, as it would require to artificially generate configurations (e.g., queue scheduling, routing) that lead to service disruption. A reasonable alternative is to create these datasets in controlled testbeds, where it is possible to use different traffic models and implement a broad set of configurations. Thus, the GNN model can be trained on samples from this testbed, and then be applied to real networks. 
Hence, the research challenge is: how to design a GNN able to provide accurate estimates in networks not seen during training? This includes topologies, traffic, and configurations (e.g., queue scheduling, routing) different from those seen in the training network testbed. 

\textbf{Generalization to larger networks:} 
From a practical standpoint, the GNN model must also generalize to \emph{larger} networks. Real-world networks include hundreds or thousands of nodes, and building a network testbed at this scale is typically unfeasible. As a result, the GNN model should be able to generalize from small network testbeds to considerably larger networks, by at least a factor of 10x.
Generalizing to larger networks \mbox{-- or} graphs, in \mbox{general --} is currently an open research challenge in the field of GNNs. 
We address this by using domain-specific knowledge from computer networks, and by proposing a novel GNN architecture that can effectively model relevant scale-independent features of networks that affect performance metrics.


\textbf{Queues and Scheduling policies:} A fundamental aspect when modeling networks is considering the behavior of queues (e.g., number, size), scheduling policies (e.g., WFQ, DRR), and the mapping of traffic flows to different Quality-of-Service classes if any. QT is a well-established technique, and models have been developed to support a wide range of scheduling policies~\cite{wfqmodel, KAPADIA1984337}. The challenge is, how to represent queues and scheduling policies inside the GNN architecture?




\vspace{0cm}
\section{Background: GNNs} \label{sec:background}
\vspace{0cm}
Graphs are used to represent relational information. Particularly, a graph $G \in \{V,E\}$ comprises a set of objects $V$ (i.e., vertices) and some relations between them $E$ (i.e., edges).

GNN~\cite{scarselli2008graph} is a family of NNs especially designed to work with graph-structured data. These models dynamically build their internal NN architecture based on the input graph. For this, they use a modular NN structure that represents explicitly the elements and connections of the graph. As a result, they support graphs of variable size and structure, and their graph processing mechanism is invariant to node and edge permutation, which eventually endows them with strong generalization capabilities over graphs -- also known as \textit{strong relational inductive bias}~\cite{battaglia2018relational}.


Despite GNN covers a broad family of neural networks with different architectural variants (e.g., \cite{scarselli2008graph, battaglia2016interaction, raposo2017discovering}), most of them share the basic principle of an iterative message-passing phase, where the different elements of the graph exchange information according to their connections, and a final readout phase uses the information encoded in graph elements to produce the final output(s).  We refer the reader to \cite{scarselli2008graph, gilmer2017neural, battaglia2018relational} for a more comprehensive background on GNNs. 


\begin{figure}[!t]
\centering
\includegraphics[width=0.9\columnwidth]{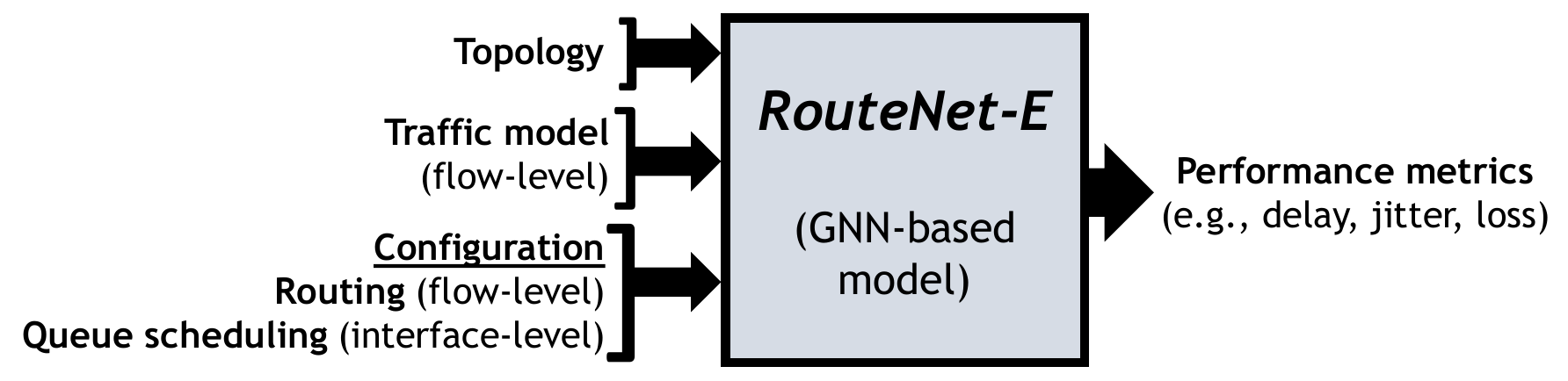}
\vspace{0cm}
\caption{Black-box representation of RouteNet-E}
\label{fig:RouteNet-E-scheme}
\vspace{-0.5cm}
\end{figure}

\vspace{0cm}
\section{RouteNet-Erlang}\label{sec:RouteNet-E}
\vspace{0cm}

This section describes RouteNet-E, a novel GNN-based solution tailored to accurately model the behavior of real network infrastructures. RouteNet-E implements a novel three-stage message passing algorithm that explicitly defines some key elements for network modeling (e.g., traffic models, queues, paths), and offers support for a wide variety of features introduced in modern networking trends (e.g., complex QoS-aware queuing policies, overlay routing).

Figure~\ref{fig:RouteNet-E-scheme} shows a black-box representation of the proposed GNN-based network model. The input of RouteNet-E is a network state sample, defined by: a network topology, a set of traffic models (flow-level), a routing scheme (flow-level), a queuing configuration (interface-level). As output, this model produces estimates of relevant performance metrics at a flow-level granularity (e.g., delay, jitter, losses). 

\vspace{0cm}
\subsection{Model Description}\label{sec:model}
\vspace{0cm}
RouteNet-E has two main building blocks: $(1)$~Finding a good representation for the network components supported by the model -- e.g., traffic models, routing, queue \mbox{scheduling --}, and $(2)$~Exploit scale-independent features of networks, in order to achieve good generalization power to larger networks than those seen during training, which is an important open challenge previously discussed in Section~\ref{sec:challenges}.

\vspace{0.1cm}
\noindent \textit{1) Representing network components and their relationships:}

First, let us define a network as a set of links \mbox{$L = \{l_i: i \in (1,...,n_l)\}$}, a set of queues on $Q = \{q_i: i \in (1,...,n_q)\}$, and a set of source-destination flows $F = \{f_i: i \in (1,...,n_f)\}$. 
According to the routing configuration, flows follow a source-destination path. Hence, we define flows as sequences with tuples of the queues and links they traverse $f_i=\{(q_{F_q(f_i,0)},l_{F_l(f_i,0)}),...,(q_{F_q(f_i,|f_i|)},$ $l_{F_l(f_i,|f_i|)})\}$, where $F_q(f_i,j)$ and $F_l(f_i,j)$ respectively return the index of the \mbox{$j$-th} queue or link along the path of flow $f_i$. Let us also define $Q_f(q_i)$ as a function that returns all the flows passing through queue $q_i$, and $L_q(l_i)$ as a function that returns the queues injecting traffic into link $l_i$ -- i.e., the queues at the output port to which the link is connected.

\begin{figure}[!t]
\centering
\includegraphics[width=0.9\columnwidth]{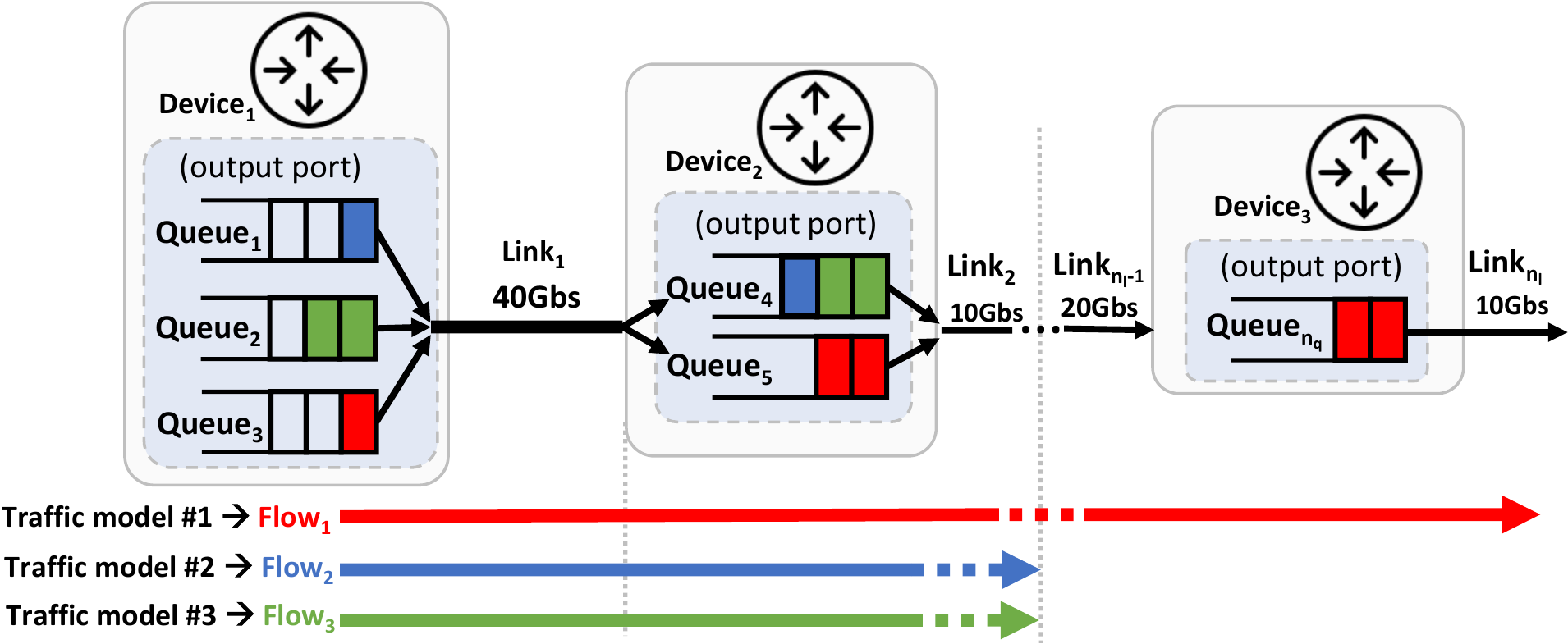}
\vspace{0cm}
\caption{Schematic representation of the network model implemented by RouteNet-E.}
\label{fig:architecture_overview}
\vspace{-0.5cm}
\end{figure}

Following the previous notation, RouteNet-E considers an input graph with three main components: $(i)$ the physical links $L$ that shape the network topology, $(ii)$ the queues $Q$ at each output port of network devices, and $(iii)$ the active flows $F$ in the network, which follow some specific src-dst paths (i.e., sequences of queues and links), and whose traffic is generated from a given traffic model. Figure~\ref{fig:architecture_overview} shows a schematic representation of the network model internally considered by RouteNet-E, which is derived from the several mechanisms that affect performance in real networks. From this model, we can extract three basic principles:

\begin{enumerate}[label=(\roman*)]
\item The state of flows (e.g., throughput, losses) is affected by the state of the queues and links they traverse (e.g., queue/link utilization).
\item The state of queues (e.g., occupation) depends on the state of the flows passing through them (e.g., traffic model).
\item The state of links (e.g., utilization) depends on the states of the queues at the output port of the link, and the 
queue scheduling policy applied over these queues.
\end{enumerate}

Formally, these principles can be formulated as follows:
\begin{align}
h_{f_k} &= g_f(h_{q_{k(0)}},h_{l_{k(0)}},...,h_{q_{k(|f_k|)}},h_{l_{k(|f_k|)}}) \label{eq:g_f}\\
h_{q_i} &= g_q(h_{p_1},...,h_{p_m}), \quad q_i \in p_k, k = 1,...,j \label{eq:g_q}\\
h_{l_j} &= g_l(h_{q_1},...,h_{q_m}), \quad q_m \in L_q(l_j) \label{eq:g_l}
\end{align}

Where $g_f$, $g_q$ and $g_l$ are some unknown functions, and $h_f$, $h_q$ and $h_l$ are latent variables that encode information about the state of flows $F$, queues $Q$, and links $L$ respectively. Note that these principles define a circular dependency between the three network components ($F$, $Q$, and $L$) that must be solved to find latent representations satisfying the equations above.

Based on the previous network modeling principles, we define the architecture of RouteNet-E (see Algorithm~\ref{alg:architecure}). Our GNN-based model implements a custom three-stage message-passing algorithm that combines the states of flows, queues and links according to Equations~\eqref{eq:g_f}-\eqref{eq:g_l}, thus aiming to resolve the circular dependencies defined in such functions. First, the hidden states $h_l$, $h_q$, and $h_f$ -- represented as n-element vectors -- are initialized with some features (lines \ref{init-l}-\ref{init-p}), denoted respectively by $x_{l_i}$, $x_{q_j}$ and $x_{f_k}$. In our case, we set the initial features of links ($x_l$) as: $(i)$ the link capacity ($C_i$), and \mbox{$(ii)$ the} scheduling policy at the output port of the link (FIFO, SP, WFQ, or DRR~\cite{shreedhar1996efficient}), using one-hot encoding. For the initial features of queues ($x_q$) we include: $(i)$ the buffer size, $(ii)$ the priority level (one-hot encoding), and $(iii)$ the weight (only for WFQ and DRR). Lastly, the initial flow features ($x_f$) are a descriptor of the traffic model used in the flow ($T_i$). Once the states are initialized, the message-passing phase is iteratively executed $T$ times (loop from line~\ref{init-loop}), where $T$ is a configurable parameter. Each message-passing iteration is in turn divided in three stages, that respectively represent the message passing and update of the hidden states of flows $h_f$ (lines~\ref{mp-path-init}-\ref{mp-path-end}), queues $h_q$ (lines~\ref{mp-queue-init}-\ref{mp-queue-end}), and links $h_l$ (lines~\ref{mp-link-init}-\ref{mp-link-end}).

\algnewcommand\algorithmicforeach{\textbf{for each}}
\algdef{S}[FOR]{ForEach}[1]{\algorithmicforeach\ #1\ \algorithmicdo}
\renewcommand{\algorithmicrequire}{\textbf{Input:}}
\renewcommand{\algorithmicensure}{\textbf{Output:}}

\begin{algorithm}[!t] \caption{Internal architecture of RouteNet-E}
\begin{algorithmic}[1]
\Require{$F$, $Q$, $L$, $x_f$, $x_q$, $x_l$}
\Ensure{$h^T_q$, $h^T_l$, $h^T_f$, $\hat{y_f}$, $\hat{y_q}$, $\hat{y_l}$}
\ForEach {$l \in L$} $h^0_l \gets [x_{l},0...0]$ \EndFor \label{init-l}
\ForEach {$q \in Q$} $h^0_q \gets [x_{q},0...0]$ \EndFor \ForEach {$f \in F$} $h^0_f \gets [x_{f},0...0]$ \EndFor \label{init-p}
\For{t = 0 to T-1} \label{init-loop} \Comment{\footnotesize Message Passing Phase}
    \ForEach {$f \in F$} \label{mp-path-init} \Comment{\footnotesize Message Passing on Flows}
        \ForEach {$(q,l) \in f$}
            \State $h^t_{f} \gets FRNN(h^t_{f},[h^t_q,h^t_l])$ \label{lin:u-flow} \Comment{\footnotesize Flow: Aggr. and Update}
            \State $\widetilde{m}^{t+1}_{f,q} \gets h^t_{f} $ \Comment{\footnotesize Flow: Message Generation}
        \EndFor
    \State $h^{t+1}_{f} \gets h^{t}_{f} $
    \EndFor \label{mp-path-end}
    \ForEach {$q \in Q$} \label{mp-queue-init} \Comment{\footnotesize Message Passing on Queues}
        \State $M_q^{t+1} \gets \sum_{f \in Q_f(q)}  \widetilde{m}^{t+1}_{f,q}$ \Comment{\footnotesize Queue: Aggregation}
        \vspace{0.1cm}
        \State $h^{t+1}_q \gets U_q(h^t_q,M_q^{t+1})$ \Comment{\footnotesize Queue: Update}
        \State $\widetilde{m}^{t+1}_{q} \gets h^{t+1}_{q} $ \Comment{\footnotesize Queue: Message Generation}
    \EndFor \label{mp-queue-end}
    \ForEach {$l \in L$} \label{mp-link-init} \Comment{\footnotesize Message Passing on Links}
        \ForEach {$q \in L_q(l)$}
            \State ${h}^{t}_{l}  \gets LRNN(h^t_l,\widetilde{m}^{t+1}_{q})$  \label{lin:u-link} \Comment{\footnotesize Link: Aggr. and Update}
	    \EndFor
	    \State $h^{t+1}_l \gets {h}^{t}_{l}$
    \EndFor \label{mp-link-end}
\EndFor \label{end-loop}
\State $\hat{y_f} \gets R_f(h^T_f)$ \label{r-path} \Comment{\footnotesize Readout phase}
\State $\hat{y_q} \gets R_q(h^T_q)$ \label{r-queue}
\end{algorithmic}
\label{alg:architecure}
\end{algorithm}
\setlength{\textfloatsep}{0.2cm}

Finally, functions $R_f$ (line~\ref{r-path}) and $R_q$ (line~\ref{r-queue}) represent independent readout functions that can be respectively applied to the hidden states of flows $h_f$ or queues $h_q$. In our experiments in Section~\ref{sec:evaluation}, we use $R_f$ and $R_q$ to predict the flow-level delay, jitter and losses -- as described later in this section.

The main motivation to use data-driven methods, such as RouteNet-E, instead of traditional QT is to achieve accurate modeling of complex traffic models that better reflect real-world traffic -- as previously introduced in Section~\ref{sec:challenges}. Hence, in RouteNet-E the representation of the traffic model descriptors ($T_i$) is central to achieve accurate modeling of different traffic patterns, and capturing their intrinsic properties. Particularly, we define $T_i$ as an n-element vector that includes the specific parameters that shape each traffic model. Find more details about the parameters of each model in section~\ref{subsec:sim-setup}.

\vspace{0.15cm}
\noindent \textit{2) Scaling to larger networks: scale-independent features}
\vspace{0.1cm}

As previously discussed in Section~\ref{sec:challenges}, generating datasets directly from networks in production would imply testing configurations that may break the correct operation. As a result, GNN-based network models should be typically trained with data from network testbeds, which are usually much smaller than real networks. In this context, it is essential for our GNN to effectively scale to larger networks than those of the training dataset -- by at least a 10x factor.

GNNs have shown an unprecedented capability to generalize over graph-structured data~\cite{battaglia2018relational,zhou2018graph}. In the context of generalizing to larger graphs, it is well known that these models keep good generalization capabilities as long as the spectral properties of graphs are similar to those seen during training~\cite{ruiz2020graphon}. In the case of RouteNet-E, its message-passing algorithm can analogously generalize to graphs with similar structures to those seen during the training phase -- e.g., similar number of queues at output ports, or similar number of flows aggregated in queues. In this vein, generating a representative dataset for RouteNet-E in small networks, covering a wide range of graph structures, does not imply any practical limitation to then achieve good generalization properties to larger networks. It can be done by simply adding a broad combination of realistic network samples with a wide variety of traffic models, routing schemes, and queuing policies as in the process described later in section~\ref{subsec:sim-setup}.

However, from a practical standpoint, scaling to larger networks often entails a broader definition beyond the topology size and structure. In particular, there are two main properties we can observe as networks become larger: $(i)$ \textit{higher link capacities} (as there is more aggregated traffic in the core links of the network), and $(ii)$ \textit{longer paths} (as the network diameter becomes larger). This requires devising mechanisms to effectively scale on these two features.

\vspace{0.1cm}
\noindent \textbf{Scaling to larger link capacities:} If we observe the internal architecture of RouteNet-E (Algorithm~\ref{alg:architecure}), we can find that the link capacity $C$ is only represented as an initial feature of links' hidden states $x_{l_i}$. The fact that $C$ is encoded as a numerical feature in the model introduces inherent limitations to scale to larger capacity values. Indeed, scaling to out-of-distribution numerical values is widely recognized as a generalized limiting factor among all neural networks~\cite{engstrom2019exploring,su2019one}. Thus, our approach is to exploit particularities from the network domain to find scale-independent representations that can define link capacities and how they relate to other link-level features that impact performance (e.g., the aggregated traffic in the link), as the final goal of RouteNet-E is to accurately estimate performance metrics (e.g., delay, jitter, losses). Inspired by traditional QT methods, we aim to encode in RouteNet-E the relative ratio between the arrival rates on links (based on the traffic aggregated in the link), and the service times (based on the link capacity), thus enabling the possibility to infer the output performance metrics of our model from scale-independent values. As a result, we define link capacities ($Cap_{link}$) as the product of a \textit{virtual reference link capacity} ($C_{ref}$) and a \textit{scale factor} ($S_{f}$) -- i.e., $Cap_{link} = C_{ref} * S_{f}$.

This representation enables to define arbitrary combinations of scale factors and reference link capacities to define the actual capacity of links in networks. Hence, in RouteNet-E we introduce the capacity feature ($C_i$) as a 2-element vector defined as $C_i$$=$$[C_{ref}, S_f]$, which is included in the initial feature vector of links ($x_l$). Note that this feature will eventually be encoded in the hidden states of links ($h_l$). In the internal architecture of RouteNet-E (Algorithm~\ref{alg:architecure}), this factor will mainly affect the update functions of flows and links (lines~\ref{lin:u-flow} and \ref{lin:u-link}), as they are the only ones that process directly the hidden states of links ($h_l$). As a result, the RNNs approximating these update functions can potentially learn to make accurate estimates on any combination of $C_{ref}$ and $S_{f}$ as long as these two features are within the range of values observed \textit{independently} for each of them during the training phase (i.e., $S_{f}\in [s_{f_{min}}, s_{f_{max}}]$ and $C_{ref}\in [C_{{ref}_{min}}, C_{{ref}_{max}}]$). Thus, we exploit this property to devise a custom data augmentation method, where we take samples from small networks with limited link capacities and generate different combinations of $C_{ref}$ and $S_{factor}$ that enable us to scale accurately to considerably larger capacities. Note that in this process, the numerical values seen by RouteNet-E ($C_{ref}$ and $S_{factor}$) are kept in the same ranges
both in the training on small networks and the posterior inference on larger networks, thus overcoming the practical limitation of out-of-distribution predictions~\cite{engstrom2019exploring,su2019one}. More details about the proposed data augmentation process are given in Sec.~\ref{subsec:training}. 

The previous mechanism enables to keep scale-independent features along with the message-passing phase of our model (loop lines \ref{init-loop}-\ref{end-loop} in Algorithm~\ref{alg:architecure}), while it is still needed to extend the scale independence to the output layer of the model. Particularly, in this paper, we use RouteNet-E to predict the flows' delay, jitter, and losses. Note that 
the distribution of these parameters can also vary for flows traversing links with higher capacities, thus leading again to out-of-distribution values. Based on the fundamentals of QT, we overcome this potential limitation by inferring delays/jitter indirectly from the occupation of queues in the network $O_{q_i}$$\in$$[0,1]$, using the $\hat{y_q}$$=$$R_q(h_q)$ function of RouteNet-E (Algorithm~\ref{alg:architecure}). Then, we infer the flow delay/jitter as a linear combination of the waiting times in queues (inferred from $O_{q_i}$) and the transmission times of the links the flow traverses. Note that a potential advantage with respect to traditional QT models is that the queue occupation estimates produced by RouteNet-E can be more accurate, especially for complex traffic models resembling real-world traffic -- as shown later in our experimental results of Section~\ref{sec:evaluation}. Likewise, for packet loss, RouteNet-E predicts directly the percentage of packets dropped with respect to the packets that were sent by the source of the flow, thus producing a bounded value $D_{f_i}$$\in$$[0,1]$, that is estimated with the $\hat{y_f}$$=$$R_f(h_f)$ function of Algorithm~\ref{alg:architecure}.

\noindent \textbf{Scaling to longer paths:} In the internal architecture of RouteNet-E, the path length only affects to the RNN function of line \ref{lin:u-flow} (Algorithm~\ref{alg:architecure}), which collects the state of queues ($h_q$) and links ($h_l$) to update flows' states ($h_f$). The main limitation here is that this RNN can typically see during training shorter link-queue sequences than those it can find then in larger networks, that can potentially have longer paths. As a result, we define $L_{max}$ as a configurable parameter of our model that defines the maximum sequence length supported by this RNN. Then, we split flows exceeding $L_{max}$ into different queue-link sequences that are independently digested by the RNN. To keep the state along with the whole flow, in case it is divided into more than one sequence, we initialize the initial state of the RNN with the output resulting from the previous sequence.

\vspace{0cm}
\subsection{Simulation Setup} \label{subsec:sim-setup}
\vspace{0cm}
To train, validate and test RouteNet-E we use as ground truth a packet-level network simulator (OMNeT++ v5.5.1~\cite{varga2001discrete}), where network samples are labeled with performance metrics, including the flows' mean delay, jitter and losses, and queue-level statistics (e.g., occupation, packet loss). To generate these datasets, for each sample we randomly select a combination of input features (traffic model, topology, and queuing configuration) according to the descriptions below:

\subsubsection{Traffic models}

Traffic is generated using five different models with increasing levels of complexity, which ranges from a basic Poisson generation process to more realistic traffic models with strong autocorrelation and heavy-tails~\cite{traffic5}. We define below the implementation details of these models (except for the well-known Poisson and Constant Bitrate, whose only configurable parameter is the traffic intensity level):


\paragraph{On-Off}
This model defines two possible states (On or Off). The lengths of On and Off periods are randomly selected $[5, 15]$ seconds. Likewise, during the On period, packets are generated using an exponential distribution. 

\paragraph{Autocorrelated exponentials}\label{par:acfexp}
This model generates autocorrelated exponentially distributed traffic staring from the following auto-regressive (AR) process: $z_{t+1}$=$az_t$$+$$\varepsilon$, $\varepsilon$$\sim$$\text{N}(0,\sigma^2)$ where $a$$\in$$(-1,1)$ controls the level of autocorrelation. The marginal distribution of $z$ is $\text{N}(0,s^2=\sigma^2/(1-a^2))$, so $z$ can be negative.
In order to construct positive inter-arrival times, $z$ is mapped by a nonlinear transformation: $\delta_t=F_E^{-1}\left(\lambda,F_N\left(0,s^2,z_t \right)\right)$, where $F_N(0,s^2,\cdot)$ and $F_E(\lambda,\cdot)$ are respectively a CDF of the normal distribution with $\mu=0$ and variance $s^2$=$[3,12]$, and an exponential distribution with parameter $\lambda$=$[40,2000]$. The first transformation changes the distribution from normal to uniform on $(0,1)$, while the second maps it into an exponential distribution.
As a result, $\delta_t$ follows an exponential distribution with autocorrelation. 
Such a model can be interpreted as a copula~\cite{nelsen1999copula}.

\paragraph{Modulated exponentials}
This model represents an alternative autocorrelated model with higher complexity for QT than the one above and is inspired by observation from~\cite{traffic2}. Particularly, the inter-arrival times are set by a hierarchical model. Inter-arrivals follow an exponential distribution ($\text{Exp}$) whose rate is controlled by the value of a hidden AR model. 
Formally, we can describe the model as $\delta_t|z_t$$\sim$$\text{Exp}(\lambda_t$=$Ae^{z_t})$, where $A$ is scaling constant and $z$ is an AR model as in the previous traffic model.

In all the previous models, average traffic rates on src-dst flows are carefully set to cover low to quite high congestion levels across different samples, where the most congested samples have $\approx$$3\%$ of packet losses.

\subsubsection{Queuing configuration}
Each forwarding device is configured with a different scheduling policy that depends on the particular scenario of our evaluation (more details in Sec.~\ref{sec:evaluation}). Overall, we use four different queue scheduling policies: First In First Out (FIFO), Strict Priority (SP), Weighted Fair Queueing (WFQ), and Deficit Round Robin (DRR)~\cite{shreedhar1996efficient}. We consider three queues per output port (except for FIFO, with only one queue), and queues have a size of 16 or 32 packets. For WFQ and DRR, we define random queue weights.

\subsubsection{Topologies}\label{subsec:topologies}
To train the GNN model we used two different real-world topologies: NSFNET (14 nodes)~\cite{hei2004wavelength}, and GEANT (24 nodes)~\cite{barreto2012fast}. Then, we validate the accuracy of RouteNet-E in GBN (17 nodes)~\cite{pedro2011performance}. 


\vspace{0cm}
\subsection{Training} \label{subsec:training}
\vspace{0cm}
We implement RouteNet-E in TensorFlow. All the datasets, the code, and the trained models are publicly available \cite{github-erlang}. 
To train the model, we use a custom data augmentation approach that, given a link capacity ($Cap_{link}$), covers a broad combination of $S_f$ and $C_{ref}$ values, in order to eventually make the model generalize over samples with larger link capacities. Particularly, given a link capacity, in some samples, we use low values of $S_f$ with higher values of $C_{ref}$, while in other samples we make it in the opposite way. As an illustrative example, if the model is trained over samples with 1Gbps links, we can represent these capacities in different samples as $Cap_{link}$=$10$*$100Mbps$$=$$1Gbps$, or $Cap_{link}$=$1$*$1Gbps$$=$$1Gbps$. Thus, after training the model should be able to make accurate inferences on samples that combine the maximum $S_f$ and $C_{ref}$ values seen during training -- i.e., $Cap_{link}$=$10$*$1Gbps$$=$$10Gbps$. In practice, this means that the model can be trained with samples with a maximum link capacity of 1 Gbps, and then scale effectively to samples with link capacities up to 10 Gbps. Note that these numbers are just illustrative, while this data augmentation method is sufficiently general to produce in the training dataset wider ranges of $S_f$ and $C_{ref}$ given a maximum link capacity. Thus, it can be potentially exploited to represent combinations leading to arbitrarily larger capacities.

\begin{figure}[!t]
\centering
\includegraphics[width=0.65\columnwidth]{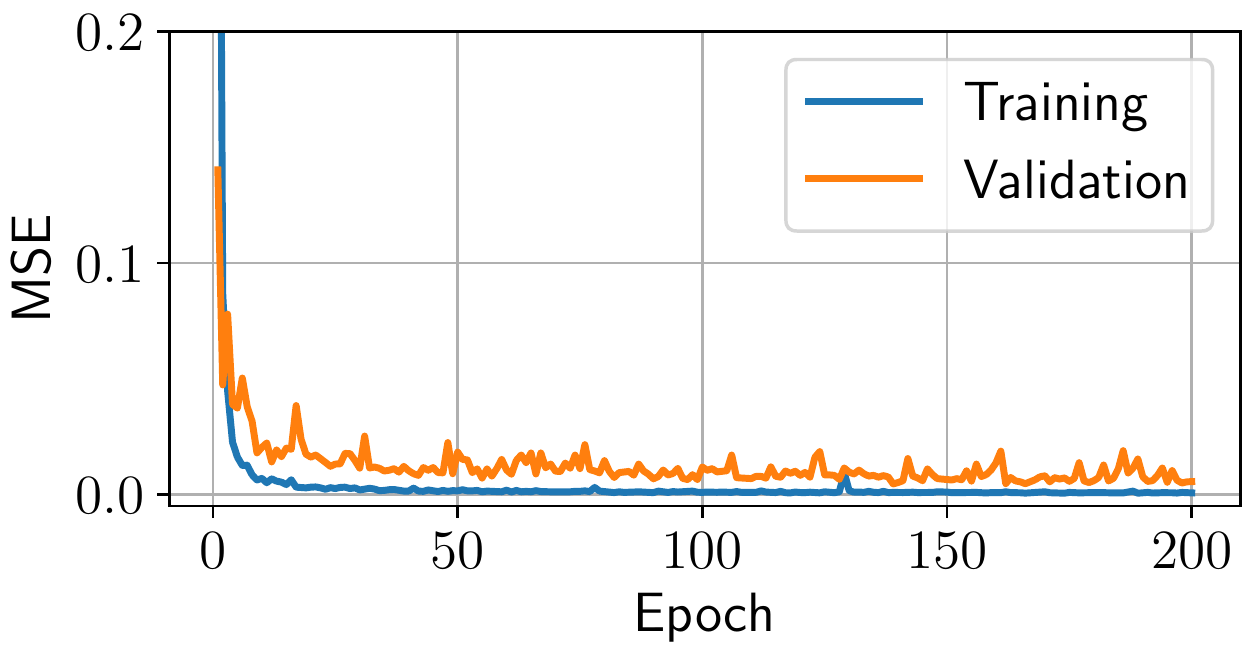}
\vspace{-0.05cm}
\caption{Training and evaluation losses over time.}
\label{fig:loss}
\vspace{0cm}
\end{figure}

After making some grid search experiments, we set a size of 32 elements for all the hidden state vectors ($h_f$, $h_q$, $h_l$), and $T$=$8$ message-passing iterations. We implement $FRNN$, $LRNN$, and $U_q$ as Gated Recurrent Units (GRU)~\cite{chung2014empirical}, and functions $R_f$ and $R_q$  as 2-layer fully-connected neural networks with ReLU activation functions. Here, it is important to note that the whole neural network architecture of RouteNet-E (Algorithm~\ref{alg:architecure}) constitutes a fully differentiable function, so it is possible to train the model end to end. Hence, all the different functions that shape its internal architecture are jointly optimized during training based on RouteNet-E's inputs (network samples) and outputs (performance metrics).

We use a training dataset with 200,000 samples from the NSFNET and GEANT topologies (100,000 samples each), including a variety of traffic model descriptors ($T_i$), routing schemes, and queue scheduling configurations -- following the descriptions in Section~\ref{subsec:sim-setup}. For the validation and test datasets, we generate 2,000 samples from the GBN topology (1,000 samples for each dataset). We train RouteNet-E for 200 epochs -- with 4,000 samples per epoch -- and set the Mean Squared Error (MSE) as loss function, using an Adam optimizer with an initial learning rate of 0.001. Figure~\ref{fig:loss} shows the evolution of the loss during training on delay estimates (for the training and validation samples), which shows stable learning along the whole training process.

\vspace{0cm}
\section{Baseline: Queuing theory}\label{qt}
\vspace{0cm}

%


In this section, we describe the state-of-the-art model we use to benchmark RouteNet-E. QT applied to networking results in a model as a function of graph-structured data.
The network is represented as a directed multigraph of queues (buffers) while edges correspond to virtual channels along with the physical link.
The general description of Equations~\eqref{eq:g_f}-\eqref{eq:g_l} holds, however, the exact relations are derived analytically from the common assumption that a system is approximated by a Markov chain. 
This makes it a perfect benchmark for RouteNet-E.

In the holistic approach, the network is modeled as a single system, like in Jackson Networks~\cite{Kelly2011} or more general BCMP queuing networks~\cite{bcmp}.
For those systems, the product form of the stationary distribution greatly simplifies the solution, however, the assumption of infinite buffers makes those models unrealistic and unable to estimate packet loss ratio.

In our approach, all the queues along the path are modeled independently.
Further, we assume that arrival to each queue is approximated by the Poisson process.
Service times are assumed to be independent and exponentially distributed.
Under those assumptions, we can derive analytical results for queue throughput, delay distribution, and blocking probability.

The aforementioned model also suffers from circular dependency.
Packet loss on a particular queue depends on its load so it also depends on the throughput of other queues feeding this particular one.
The throughput, however, depends on packet loss so we end up with circular dependence.
We fixed this problem by a \emph{map-reduce} inspired algorithm that also emphasizes the resemblance between GNN and QT.

The algorithm consists of there functions: \emph{map\_queues}, \emph{map\_paths} and \emph{reduce}.
The \emph{map\_queues} function updates packet loss for each queue, given the total traffic (external demands plus within network transfer).
The function also computes the remaining QoS parameters (jitter and delay).
The \emph{map\_paths} function updates the traffic knowing the packet loss on every queue.
Finally, the \emph{reduce} function aggregates per path delay, jitter, and packet loss.
In the first iteration, we assume no packet loss.
Given the first approximation, we can compute the loss probability (\emph{map\_queues}) and update the intensities to account for the losses (\emph{map\_paths}). 
After a few iterations, the algorithm converges to a fixed point and the final values are reduced (\emph{reduce}). 
Notice how this approach is similar to RouteNet-E forward pass.
In QT, we use known analytical relations while in GNN those relations are approximated by a neural network and learned from the data. 

For an $M/M/1/b$ system, we used known formulas for blocking probabilities and delay distribution to get average delay and jitter.
For a network with scheduling, we designed \emph{map\_queues} functions based on the Markov chain model described below.
Because scheduling couples the queues, the corresponding \emph{map\_queues} operates on groups of queues assigned to the same link. 

Let us begin with a strict priority scheduler.
Consider $p$ priority class customers arriving at rate $\lambda_i$ and requiring service time with mean $1/\mu_i$. 
Each class waits in the independent virtual queue limited by $b_i$ and served in non-preemptive FIFO order.
Such a system can be modeled as a continuous-time Markov chain on the state space $\mathcal{S}_{SP}=\{(s,q=(q_1,q_2,\ldots,q_p)\}$, where $s$ denotes the priority class currently being served or $0$ if the system is empty.
The remaining part of the state: $p$-tuple $q$ encodes the number of customers for each priority.
For convenience let us define $q_{i+}\vcentcolon=(q_1,\ldots,q_i+1,\ldots,q_p)$,
$q_{i-}\vcentcolon=(q_1,\ldots,q_i-1,\ldots,q_p)$ and $q^0=(0,\ldots,0)$.
The model is based on~\cite{KAPADIA1984337} and modified to allow for per-priority class buffer size.
Time evolution of the CTMC is characterized by the generator matrix $\bm Q$ whose elements follow the rules:
\vspace{-0.05cm}
\begin{align}
    \bm Q&[(0,q),(i,q_{i+})]=\lambda_i\quad 0< i \leq p \label{eq:Qspa}\\
    \bm Q&[(s,q),(s,q_{i+})]=\lambda_iI_{q_i<b_i}\quad 0< i \leq p \\
    \bm Q&[(s,q^0_{s+}), (0,q^0)]=\mu_s \quad 0< s \leq p\\
    \bm Q&[(s,q),(\min\{i:q_i>0\},q_{s-})]=\mu_s\label{eq:Qspz}
\end{align}
where $I_{A}$ is an indicator function and $\bm Q[.,.]$ denotes entry in generator matrix.
If neither rule matches states pair a general rules $\bm Q[s,s']=0, \quad s\neq s'$ and $\bm Q[s,s]=-\sum_{s'\neq s}\bm Q[s,s']$ apply.
A similar model can be constructed for WFQ and DRR.

Since both scheduling policies approximate an ideal Generalized Processor Sharing the same model is used for WFQ and DRR.
The CTMC is similar to \eqref{eq:Qspa}-\eqref{eq:Qspz} with exception that the queue $i$ is served at rate $\mu_i$ if other queues are empty, otherwise the rate scales proportionally to the positive weight $w_i$.
State space $\mathcal{S}_{GPS}$ is also simplified and it is formed solely of $p$ tuples $q$ defined as for SP.
The resulting CTMC is based on~\cite{wfqmodel} and evolves according to the following generator:\\
\vspace{-0.4cm}
\begin{align}
    \bm Q&[q,q_{i+}]=\lambda_i I_{q_i<b_i}\quad 0< i \leq p\\
    \bm Q&[q,q_{i-}]=\frac{I_{q_i>0} w_i}{\sum_{q_i>0} w_i}\mu_i\quad 0< i \leq p
\end{align}
Given the generator matrix $\bm Q$, we can develop either an analytical solution for queue characteristics as in~\cite{wfqmodel,KAPADIA1984337} or use a direct approach and obtain them numerically.
We chose the latter and compute packet loss delay and jitter assuming the CTMC has reached stationary distribution $\bm \pi$ computed from global balance equations (GBE)~\cite{Kelly2011} that form a sparse linear system.
%
We obtained $\bm \pi$ from sparse eigenvalue decomposition via Arnoldi method~\cite{Stewart2000} with a general sparse linear solver as a fallback in case of numerical instabilities.
Given the $\bm \pi$, the packet loss ratio for class $i$ ($\bm p_{b}[i]$) is the sum of all state probabilities where queue $i$ is full.
The delay is computed from average queue size (with respect to $\bm \pi$) using Little's law.
Computation of jitter requires a more sophisticated approach.
We pose this as the first passage time problem in CTMC~\cite{Kijima97a}.
The delay of a class $i$ customer is the first passage time to any state where the queue $i$ is empty provided that no new customers can arrive so  $\lambda_i=0$ for GPS or $\lambda_j=0, j\leq i$ for SP. 
Its conditional distribution can be calculated from $\bm Q$ using Laplace transform~\cite{Kijima97a}.
The final delay distribution and jitter are obtained from the total probability theorem.
It is assumed that a packet of class $i$ observing state $s$ at his arrival experiences delay equal to the first passage time from the state just after his arrival $s_{i+}$. 
From PASTA property, the probability of such an event is $\bm \pi[s]/({1-\bm p_{b}[i]})$, here we condition of the event that packet is not dropped.

\vspace{0cm}
\section{Evaluation}\label{sec:evaluation}
\vspace{0cm}

In this section, we evaluate the performance of RouteNet-E in a wide range of relevant scenarios. We seek to understand:\\
1) Can RouteNet-E model complex traffic models? What is the accuracy with realistic models with strong autocorrelation and heavy-tails?\\
2) Is RouteNet-E able to understand more complex multi-queue scheduling policies? What is the accuracy compared to QT?\\
3) Is RouteNet-E able to generalize to unseen network configurations and traffic loads? Also, can it generalize to \emph{larger} networks?\\
4) How fast is RouteNet-E compared to the QT benchmark? Does it allow for real-time operation?

\begin{figure*}
    \begin{subfigure}{0.30\textwidth}
    \centering
    \includegraphics[width=\textwidth]{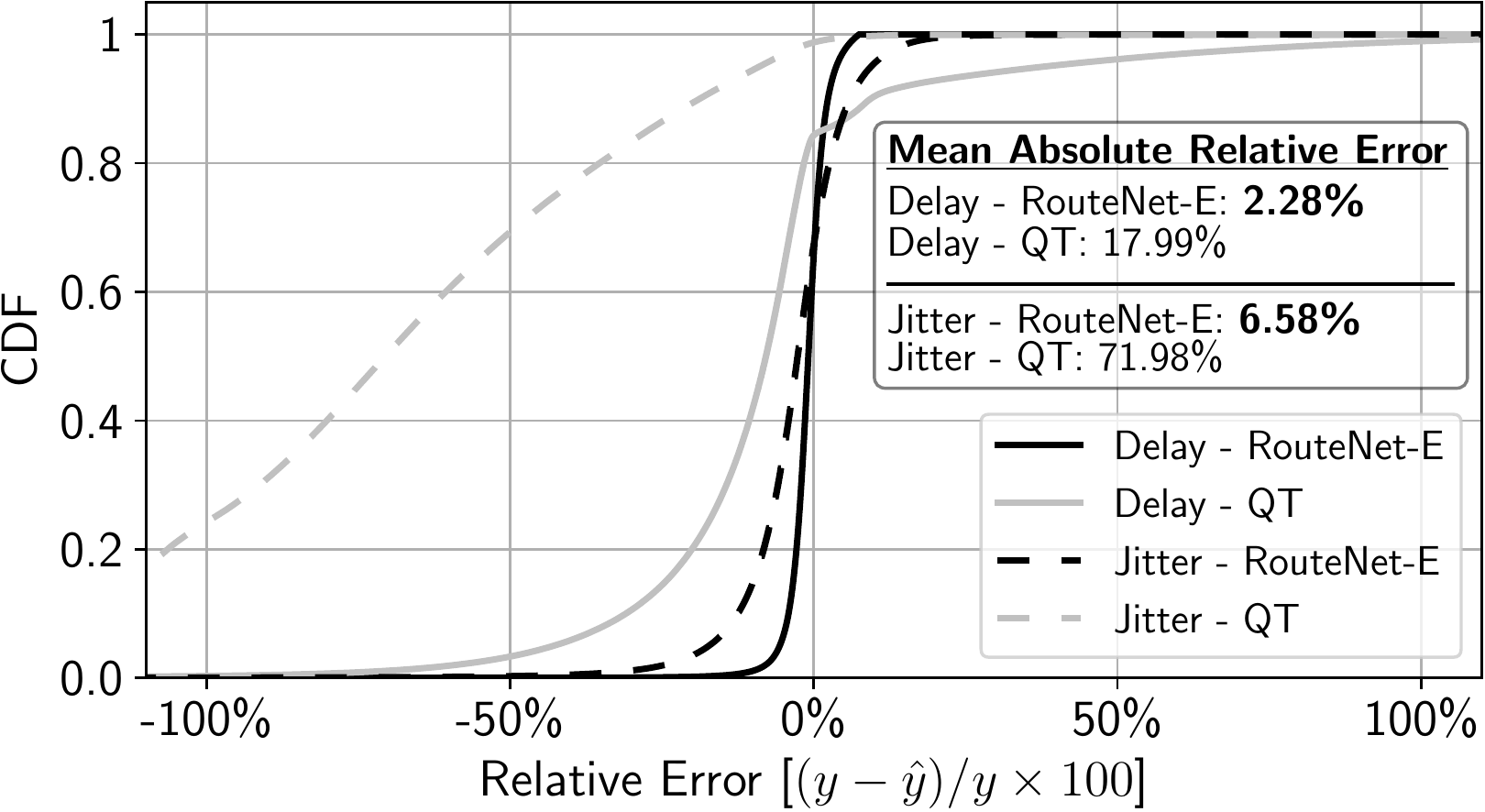}
    \caption{Poisson}
    \label{fig:poisson}
    \end{subfigure}\hfill
    \begin{subfigure}{0.30\textwidth}
    \centering
    \includegraphics[width=\textwidth]{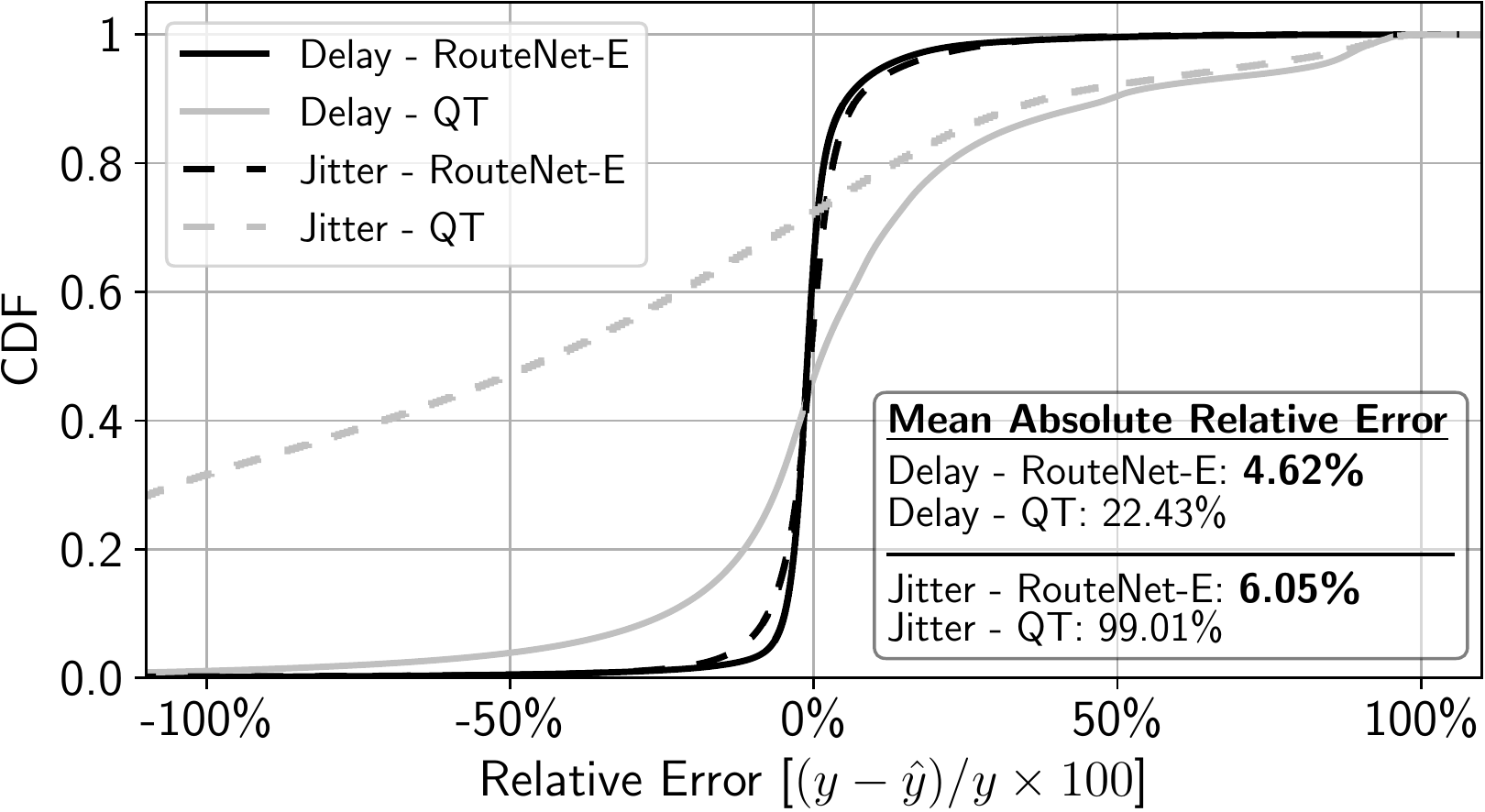}
    \caption{Constant bitrate}
    \label{fig:periodic}
    \end{subfigure}\hfill
    \begin{subfigure}{0.30\textwidth}
    \centering
    \includegraphics[width=\textwidth]{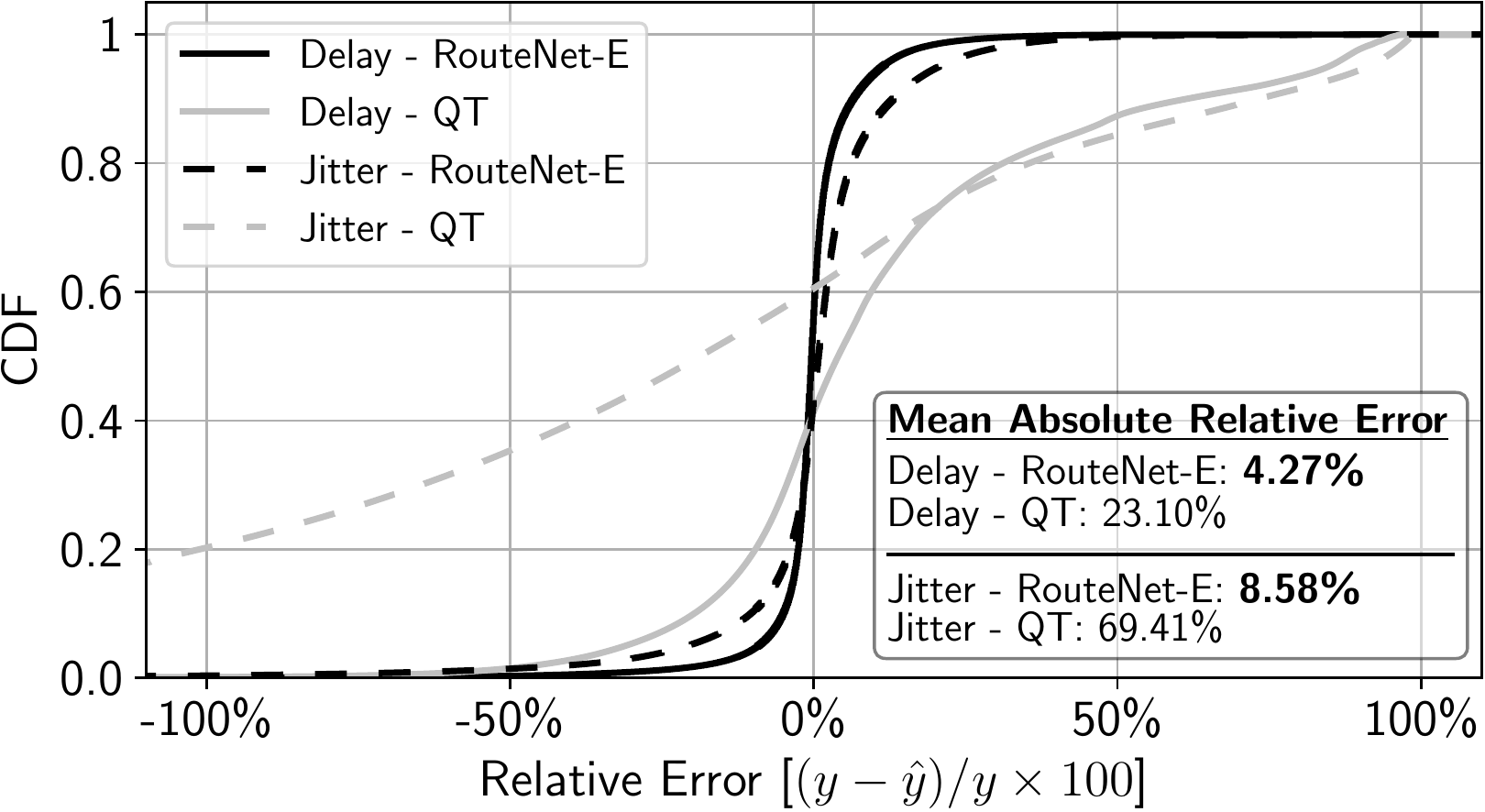}
    \caption{On-Off}
    \label{fig:onoff}
    \end{subfigure}

\medskip
    \begin{subfigure}{0.30\textwidth}
    \centering
    \includegraphics[width=\textwidth]{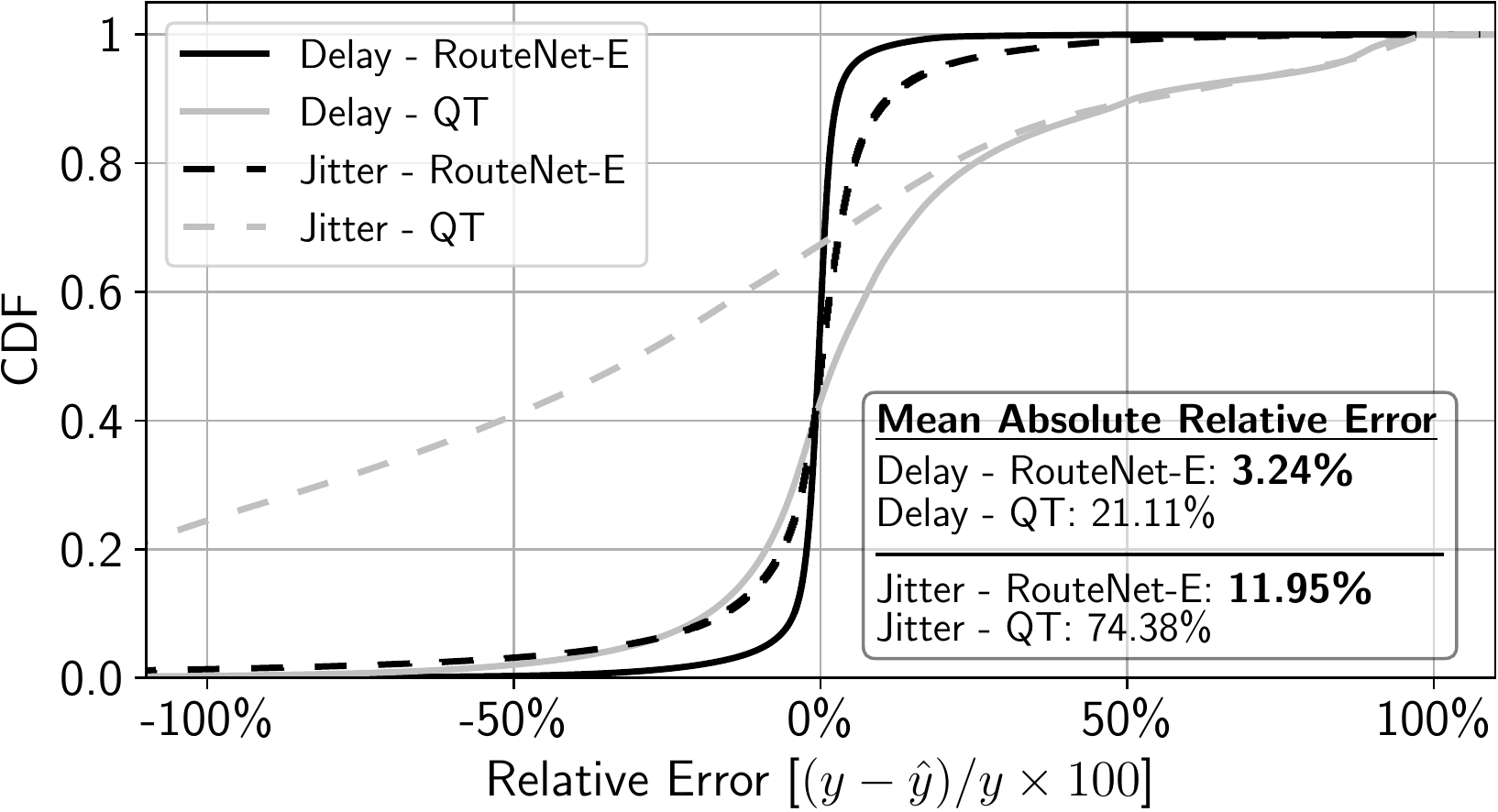}
    \caption{Autocorrelated exponentials}
    \label{fig:auto_exp}
    \end{subfigure}\hfill
    \begin{subfigure}{0.30\textwidth}
    \centering
    \includegraphics[width=\textwidth]{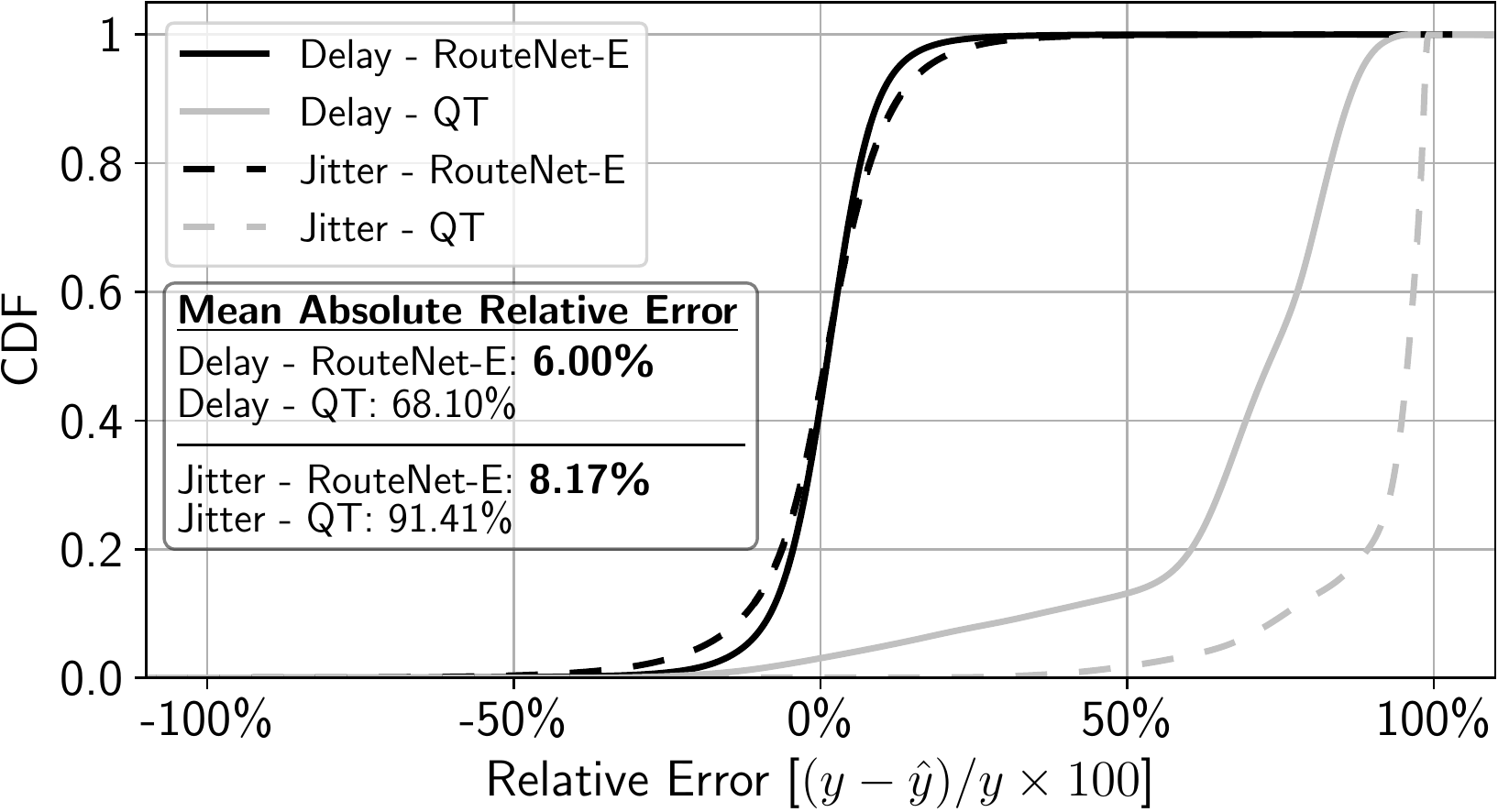}
    \caption{Modulated exponentials}
    \label{fig:mod_exp}
   \end{subfigure}\hfill
    \begin{subfigure}{0.30\textwidth}
    \centering
    \includegraphics[width=\textwidth]{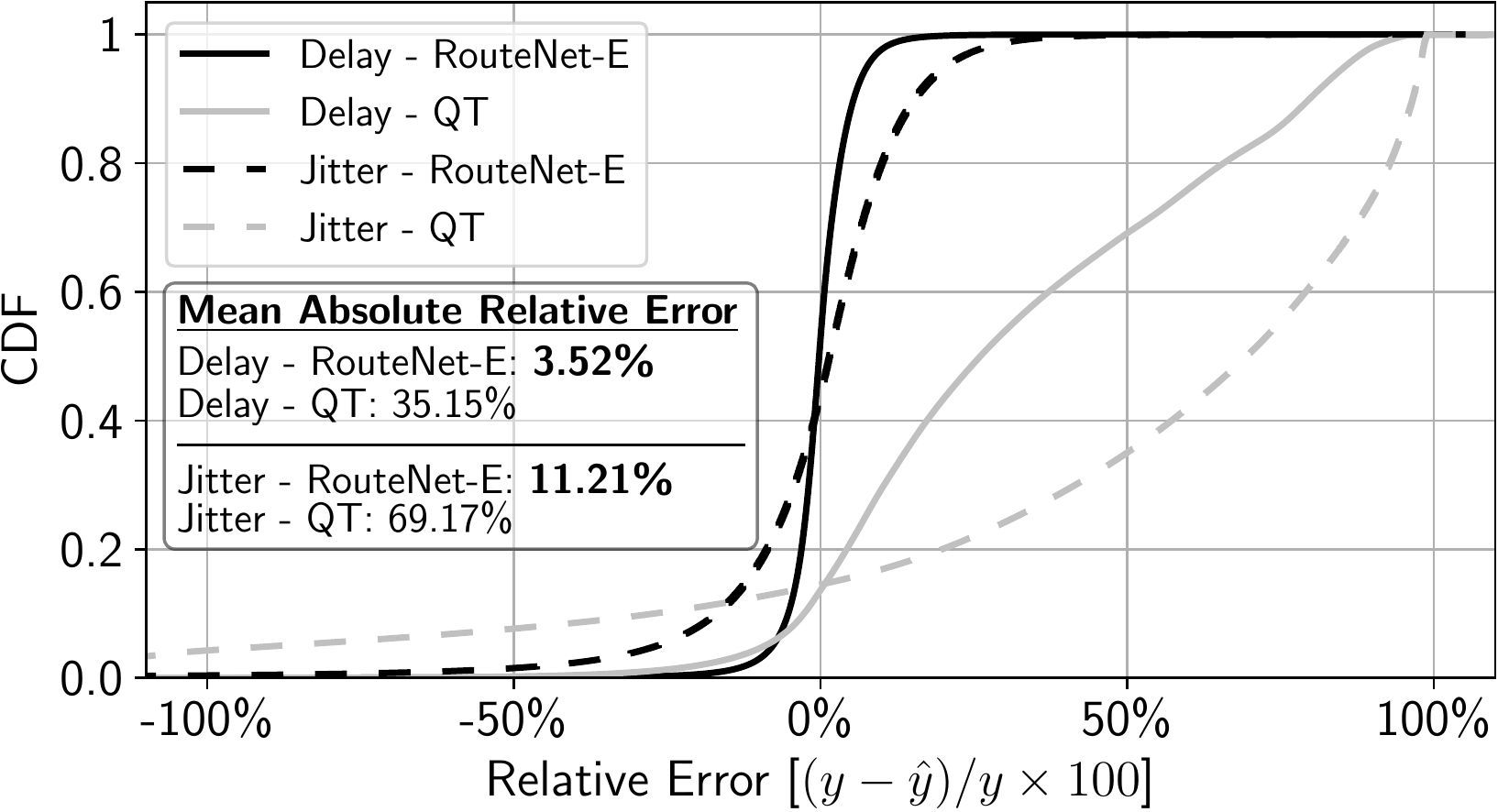}
    \caption{All traffic models multiplexed}
    \label{fig:all}
    \end{subfigure}

\medskip
\vspace{-0.05cm}
\caption{CDF of the relative error for RouteNet-E and QT with different traffic models. Top row shows models with discrete state space, bottom row includes continuous state space. Each figure also shows numbers of the mean absolute relative error.}
\label{fig:traffic_model}
\vspace{-0.4cm}

\end{figure*}

\subsection{Evaluation Methodology}
\vspace{0cm}
To analyze the accuracy of RouteNet-E (Sec.~\ref{sec:RouteNet-E}) and benchmark it against the state-of-the-art queuing theory model (Sec.~\ref{qt}), we use the following methodology. In all the experiments the ground-truth is obtained with a packet-level simulator (see Sec.~\ref{subsec:sim-setup} for details). Unless noted otherwise, in each evaluation we perform 50k experiments with a random configuration (src-dst routing, traffic intensity, per-interface scheduling policy, and queue length) and compute the mean average delay, jitter, and losses. Then, we compute the error of RouteNet-E's and QT's estimates. For a fair comparison, we use samples of the GBN topology, which is not included during training (see Sec.~\ref{subsec:training} for training details). Finally, depending on the experiment we use different traffic models (Sec.~\ref{subsec:sim-setup}) and a wide range of realistic topologies. 

\vspace{0cm}
\subsection{Traffic Models}\label{sec:traffic_models}
This section focuses on analyzing the accuracy of RouteNet-E in a wide range of traffic models. The experiment is organized such that we add complexity to the traffic model by changing its first and second-order statistics (i.e., variance and autocorrelation). With this, we use challenging models that are good approximations to those seen in Internet links.

Figure \ref{fig:traffic_model} shows the CDF of the relative error (in \%) for all the traffic models under evaluation. We plot the error for the delay and jitter estimates of both, RouteNet-E and QT. As we can observe, RouteNet-E achieves excellent results, producing very accurate estimates of delay and jitter in all traffic models, with a worst-case error below $6\%$ for delay and $12\%$ for jitter (mean absolute relative error). 

As expected, QT results in unacceptable performance in continuous-state traffic models (up to $68\%$ for delay), while it achieves moderate accuracy for discrete-state models. Interestingly, QT shows poor accuracy across all the experiments estimating jitter. The reason for this is that QT assumes independence between queues in the network. Hence, the estimator used for jitter is the sum of the individual delay variance of queues along flow paths, which ignores possible covariance effects between queues.

It is remarkable that RouteNet-E is also accurate even with non-Markovian traffic models (On-Off, Figure~\ref{fig:onoff}) and with challenging models that approximate strong autocorrelation (Autocorrelated Exponentials, Figure \ref{fig:auto_exp}). For the latter, it has been shown in the literature that the TCP protocol generates traffic with autocorrelation for a finite range of time-scales~\cite{figueiredo2002autocorrelation}. In this scenario, RouteNet-E estimates the delay with a mean error of $3.24\%$.

Figure \ref{fig:mod_exp} plots the accuracy for the Modulated Exponentials model, this emulates observations found at Internet links~\cite{traffic5} by approximating a heavy-tail. In this scenario, RouteNet-E still produces very accurate estimates. It is worth noting that this traffic model could be made even more difficult for QT by increasing both the variance and the autocorrelation factor.

The key to RouteNet-E's performance is that it has been trained for such traffic models. As discussed in Section~\ref{sec:RouteNet-E}, we have parameterized the models and trained the GNN to learn the interaction between the traffic, the queues, and the resulting performance metrics. The experiments depicted in Figure \ref{fig:traffic_model} show that RouteNet-E can generalize to traffic, providing good accuracy even for traffic models with parameters not seen in training. RouteNet-E is designed to be an extensive model, adding a new traffic model is as simple as pasteurizing it and including it in training.

To showcase this, consider the experiment shown in Figure~\ref{fig:all}, where we run 100k experiments with samples where each src-dst pair uses a \emph{random traffic model} with random parameters. Effectively, we multiplex all traffic models in a single network topology. As the figure shows, RouteNet-E is able to model this scenario in the presence of complex interactions of various multiplexed traffic models.

\vspace{0cm}
\subsection{Scheduling Policies} \label{subsec:sched_policies}

With this experiment, we aim to validate that RouteNet-E is able to model the behavior of queues. For this we use 100K samples of the GBN topology, each router port is configured with three different queues and with a randomly selected scheduling policy (FIFO, WFQ, DRR, SP). For WFQ and DRR, the set of weights is also randomly assigned. Moreover, each src-dst path is assigned a Quality-of-Service class that maps traffic flows to specific queues. In order to provide a fair benchmark with QT, we use only Poisson traffic.

Table \ref{tab:sched} summarizes the results, which are grouped for various traffic intensities, from low-loaded to highly-congested scenarios, where the mean packet loss rate is around $3\%$. As we can observe, RouteNet-E outperforms QT, obtaining highly accurate estimates for all the evaluated metrics.

\begin{table}[!t]
\setlength{\tabcolsep}{2pt}

\resizebox{1\columnwidth}{!}{
\centering
\begin{tabular}{c|ccc|ccc|ccc|}
    & \multicolumn{3}{c|}{\textbf{Delay}}  & \multicolumn{3}{c|}{\textbf{Jitter}}  & \multicolumn{3}{c|}{\textbf{Loss}} \\ \cline{2-10} 
    & \textbf{Low}     & \textbf{Med}  & \textbf{High}    & \textbf{Low}     & \textbf{Med}  & \textbf{High}     & \textbf{Low}     & \textbf{Med}  & \textbf{High}    \\ \cline{2-10} 
\textbf{RouteNet-E} & $2.0\%$ & $2.2\%$ & $3.3\%$ & $4.8\%$ & $6.2\%$ & $10.6\%$ & $12.61\%$ & $12.7\%$ & $12.66\%$ \\ 
\textbf{QT}  &   $13.0\%$      &   $17.3\%$      &  $25.1\%$       &    $49.0\%$     &   $53.2\%$      &    $59.6\%$      &  $61.83\%$       &   $59.3\%$      &   $57.9\%$ 

\end{tabular}%
}
\caption{Results for Scheduling Policies}
\label{tab:sched}
\vspace{-0.05cm}
\end{table}

\vspace{0cm}
\subsection{Generalization to larger topologies}\label{sec:size_gen}

The previous experiments have shown that RouteNet-E achieves remarkable accuracy in performance evaluation under different traffic models (Sec.~\ref{sec:traffic_models}) as well as complex scheduling policies (Sec.~\ref{subsec:sched_policies}). As we have discussed in Section~\ref{sec:challenges}, ML-based network models must generalize to unseen and \emph{larger} networks to become a practical solution. In this vein, RouteNet-E was carefully designed to address this challenge (see Sec.~\ref{sec:model} for details). 

In this set of experiments, we evaluate RouteNet-E in a wide range of networks considerably larger than the ones seen during training. Specifically, the model has been trained with topologies between 25 and 50 nodes and tested with topologies from 50 to 300 nodes. All these networks have been artificially generated using the Power-Law Out-Degree algorithm described in~\cite{palmer2000generating}, where the ranges of the $\alpha$ and $\beta$ parameters have been extrapolated from real-world topologies of the Internet Topology Zoo repository~\cite{6027859}. Link capacities and 
the generated traffic volume is scaled accordingly.



Figure~\ref{fig:scalability} shows how RouteNet-E generalizes to larger topologies not seen in training. Specifically, the boxplots show the absolute relative error with respect to the topology size. As expected, RouteNet-E obtains better accuracy in topologies that are closer to the ones seen during the training phase (50 to 99 nodes), achieving an average error of $4.5\%$ (green line). As the topology size increases, the average error stabilizes to $\approx 10\%$. Note that this value is even lower than the one obtained by the QT model, which achieves a mean error of $12.6\%$ in samples with Poisson traffic (Fig.~\ref{fig:poisson}). We could not test larger topologies ($>$300) in our cluster (180 nodes), as packet-level simulations -- used for the ground-truth -- are sequential in nature, and, with our traffic configurations, have exponential complexity with respect to the topology size.

Generalization is an open challenge in the field of GNN. As discussed in Sec.~\ref{sec:RouteNet-E}, we have addressed this by using domain-specific knowledge and data augmentation. Particularly, we infer delay/jitter from queues' occupation and apply our scale-independent method to generalize to larger topologies.

\begin{figure}[!t]
\centering
\begin{minipage}[t]{.49\columnwidth}
  \centering
    \vspace{0cm}
  \includegraphics[width=\linewidth]{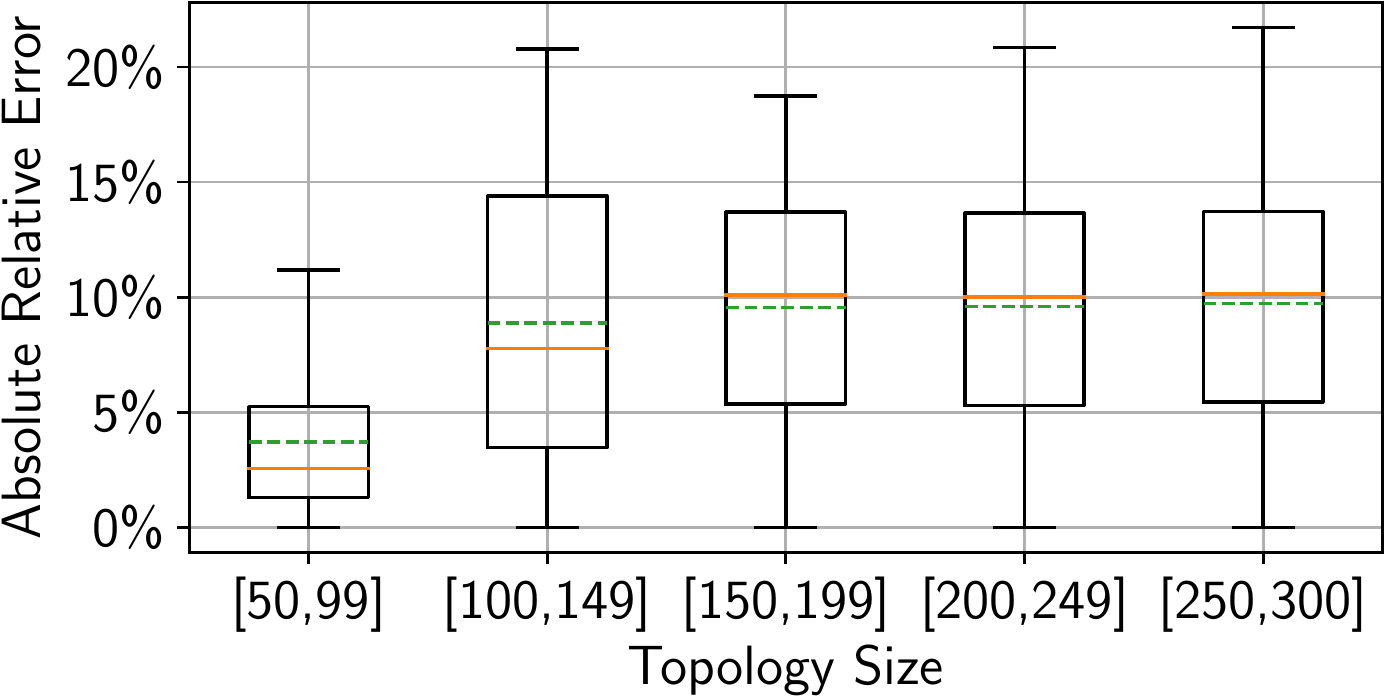}
  \vspace{-0.4cm}
  \captionsetup{width=0.8\linewidth}
  \captionof{figure}{Absolute relative error vs. topology size.}
  \label{fig:scalability}
  \vspace{-0.5cm}
\end{minipage}%
\hfill
\begin{minipage}[t]{.49\columnwidth}
  \centering
    \vspace{0cm}
  \includegraphics[width=\linewidth]{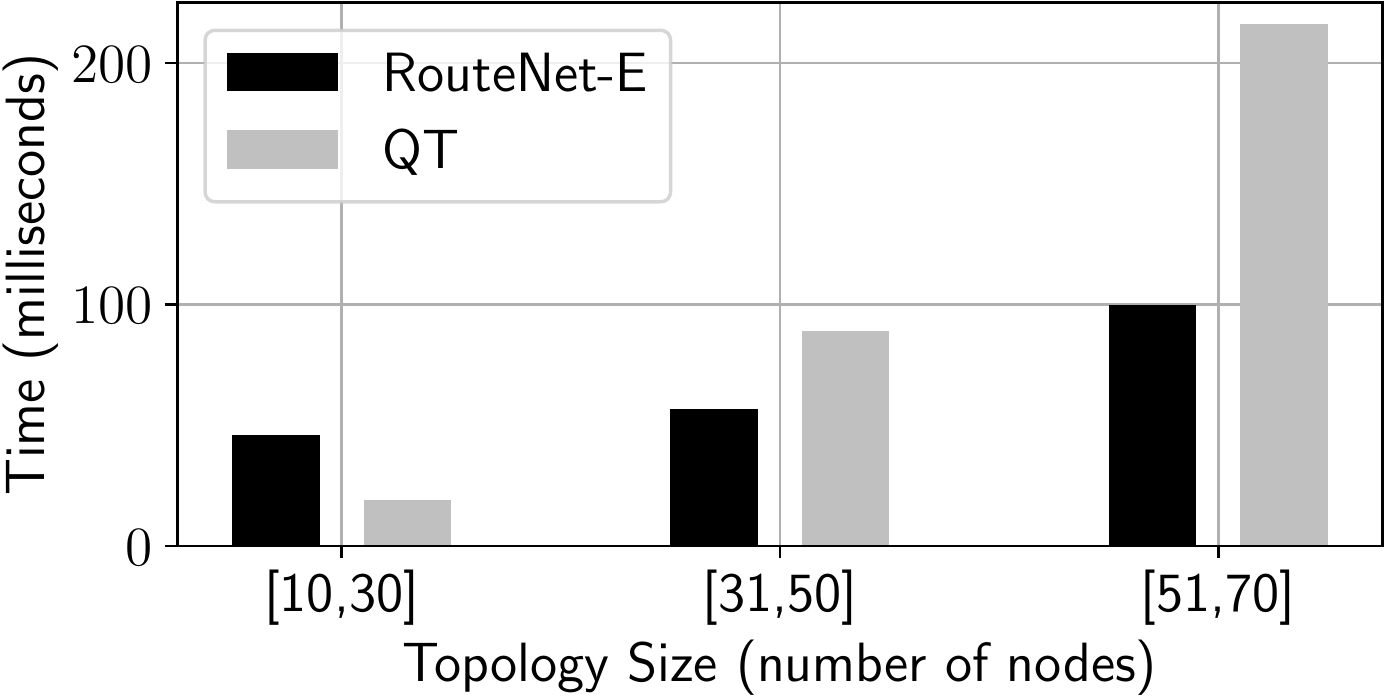}
  \vspace{-0.4cm}
  \captionsetup{width=0.8\linewidth}
  \captionof{figure}{Execution time vs. topology size.}
  \label{fig:exec_time}
  \vspace{-0.05cm}
\end{minipage}
\end{figure}

\vspace{0cm}
\subsection{Inference Speed}

Finally, in this section, we evaluate the inference speed of RouteNet-E. Fast models are especially appealing for network control and management, as they can be deployed in real-time scenarios. For this, we have measured the execution times [Intel(R) Xeon(R) Gold 5220 CPU @ 2.20GHz] of the experiments in the previous section, for both QT and RouteNet-E. The results (Figure~\ref{fig:exec_time}) show that both models operate in the order of milliseconds. In particular, RouteNet-E goes from a few milliseconds for small topologies to a few hundred for the larger ones.

\vspace{0cm}
\section{Related Work}
\vspace{0cm}

The use of Deep Learning (DL) for network modeling has recently attracted the interest of the networking community. This idea was initially suggested by \emph{Wang, et al.} \cite{wang2017machine}. The authors survey different techniques and discuss data-driven models that can learn real networks. Initial attempts to instantiate this idea use fully connected neural networks (e.g. \cite{valadarsky2017learning, mestres2018understanding}). Such early attempts do not generalize to networks not seen in training, are not tested with realistic traffic models, and do not model queues. More recent works propose more elaborated neural network models, like Variational Auto-encoders \cite{xiao2018deep} or ConvNN \cite{chen2018deep}. However, they have similar limitations. 

Finally, some early pioneering works use GNN in the field of computer networks \cite{geyer2019deeptma, rusek2019unveiling, meng2020interpreting}. However, they use a basic GNN architecture that considers a simplified model of the network, 
ignoring traffic models, queuing policies, and the critical property of generalizing to larger networks.


\vspace{0cm}
\section{Discussion and Concluding remarks}
\vspace{0cm}

In this paper, we have presented RouteNet-E, a new tool for network modeling. RouteNet-E has shown remarkable accuracy in all the scenarios, outperforming a state-of-the-art QT model. RouteNet-E also overcomes the main limitation of QT, and it is able to model challenging traffic models. More importantly, the proposed model addresses the main drawback of existing ML-based models, and it is able to provide accurate estimates in larger networks ($\approx$10x).

RouteNet-E provides unprecedented accuracy in network performance evaluation. However, in contrast to QT, it does not help understand the behavior of the network being modeled. The knowledge learned by RouteNet-E during training is not human-understandable. This is a common issue for all ML-based models, and substantial research efforts are being devoted to producing explainable ML models \cite{meng2020interpreting}. However, this is still an open research problem.

RouteNet-E's performance enables network optimization, planning, and operation in real-time scenarios. It also represents an open-source extensible model. We hope that the community will use it as a baseline to incorporate additional network components, such as other scheduling policies, traffic models, etc. 


\bibliographystyle{IEEEtran}
\bibliography{bibliography}

\end{document}